%% file: main.tex
\documentclass[journal]{vgtc}                     

\onlineid{1169}

\vgtccategory{Analytics \& Decisions}

\title{HiLSVA: Design and Evaluation of a Human-in-the-Loop \\ Agentic System for Scientific Visualization}

\author{%
  \authororcid{Kuangshi Ai}{0009-0005-7171-6529},
  \authororcid{Patrick Phuoc Do}{0009-0001-8966-6823}, and 
  \authororcid{Chaoli Wang}{0000-0002-0859-3619}
}

\authorfooter{
  \item The authors are with 
the Department of Computer Science and Engineering, University of Notre Dame, Notre Dame, IN 46556, USA.\\
E-mail: \{kai, mdo23, chaoli.wang\}@nd.edu.
}

\abstract{%
Large language model (LLM) agents enable natural language interaction for scientific visualization (SciVis). Still, prior systems have essentially prioritized autonomy over human analytical control, thereby limiting transparency and \hot{human oversight}. We present HiLSVA, a human-in-the-loop agentic system that supports mixed-initiative SciVis workflows. HiLSVA integrates a plan-first multi-agent architecture with explicit human oversight, stepwise provenance tracking, and learn-at-test-time adaptation from user feedback. The system supports fluid handoff between humans and agents through both natural language and direct manipulation of visualizations, while sandboxed execution ensures safe, reproducible workflows. In doing so, HiLSVA reframes agentic SciVis as a collaborative process that augments, rather than replaces, human analytical reasoning. We evaluate HiLSVA through representative case studies and a controlled user study with twelve participants of varying expertise across multiple autonomy settings. Results show that mixed-initiative interaction improves task completion, \hot{user control}, and workflow transparency across different levels of user expertise, while revealing a tradeoff between execution efficiency and human oversight. These findings highlight the importance of human-centered design in agentic SciVis and guide the development of future collaborative visualization systems. \hot{We encourage readers to explore our demo video, case studies, and source code at \url{https://hilsva.github.io/}.}
}


\graphicspath{{figs/}{figures/}{pictures/}{images/}{./}} 

\usepackage{tabu}                      
\usepackage{booktabs}                  
\usepackage{lipsum}                    
\usepackage{mwe}                       

\usepackage{mathptmx}                  
\usepackage{amsmath}
\usepackage{amssymb}
\usepackage{bm}
\usepackage{comment}
\usepackage{adjustbox}
\usepackage{makecell}
\usepackage{booktabs}
\usepackage{pifont}
\usepackage{graphicx}
\usepackage{multirow}

\newcommand{\hot}[1]{#1}

\begin{document}


\firstsection{Introduction}
\maketitle
\input{src/intro}
	
\vspace{-0.075in}
\input{src/related}

\vspace{-0.075in}
\input{src/method}
	
\vspace{-0.075in}
\input{src/results}

\vspace{-0.075in}
\input{src/limitations}

\vspace{-0.075in}
\input{src/conclusion}

\vspace{-0.075in}
\acknowledgments{This research was supported in part by the U.S.\ National Science Foundation through grants IIS-2101696, OAC-2104158, IIS-2401144, and IIS-2550610, and the U.S.\ Department of Energy through grant DE-SC0023145. The authors thank the anonymous reviewers for their insightful comments.}

\vspace{-0.075in}
\bibliographystyle{abbrv-doi-hyperref-narrow}
\bibliography{refs-abbv}

\appendix 
\crefalias{section}{appendix} 

\input{src/appendix}

\end{document}

%% file: src/intro.tex
Advances in large language models (LLMs) have made complex human–computer interactions more intuitive and accessible. They have reshaped general computer use by enabling conversational agents to assist with everyday tasks such as file management, email drafting, and troubleshooting~\cite{zhang2025ufo2, anthropic2024computeruse, hu-etal-2025-os, wu2024oscopilot}. In web navigation, LLM agents can search, filter, and summarize online content, assisting users in making decisions or even completing tasks such as booking tickets on their behalf~\cite{lai2025autowebglm, liu2025wepo}.
Beyond these broad applications, LLMs are transforming domains that once demanded specialized expertise. In software engineering, they can generate code, explain concepts, and aid debugging, lowering barriers for both novices and experts~\cite{liu2024large, jin2024llms}. In visualization, they bridge the gap between analytical intent and implementation, allowing users to state goals in natural language and obtain executable workflows~\cite{zhao2024lightva, liu2024ava}. Building on this trajectory, autonomous LLM-driven agents now enable natural-language instructions to be translated directly into sophisticated visualizations.

In scientific visualization (SciVis), several recent systems exemplify this paradigm as we witness the shift from generative AI~\cite{Wang-TVCG23} to agentic AI. ChatVis~\cite{peterka2025chatvis} introduces an iterative LLM agent that generates and refines ParaView Python scripts through retrieval-augmented prompting and error-driven feedback, thereby improving the robustness of code generation. ParaView-MCP~\cite{liu2025paraview} adopts the {\em model context protocol} (MCP) to directly connect multimodal LLMs with ParaView's Python API, enabling seamless collaboration among the user, agent, and visualization environment. Beyond direct tool control, NLI4VolVis~\cite{ai2025nli4volvis} advances natural language interaction for volumetric data by combining LLM multi-agents, editable 3D Gaussian splatting~\cite{Tang-PVIS25}, and semantic segmentation, supporting open-vocabulary querying and real-time scene editing. VizGenie~\cite{biswas2025vizgenie} introduces a self-improving framework that dynamically generates, validates, and integrates new visualization modules, enabling adaptive, feature-based exploration. 
These systems illustrate an emerging class of \textbf{LLM-based SciVis agents}, which can be broadly defined as {\em AI agents that interpret user intent expressed in natural language and autonomously interact with visualization pipelines to produce outputs aligned with user-specified analysis goals}~\cite{dhanoa2025agentic, Ai-GenAI25}. 

However, existing works tend to emphasize autonomy and task completion, often sidelining the central role of human analysts. As Dhanoa et al.\ \cite{dhanoa2025agentic} argued, visualization is fundamentally human-centered, emphasizing reasoning, interpretability, and accountability rather than automation alone. The critical challenge is not simply to build more autonomous SciVis agents, but to balance human analytical control, encompassing how analysts delegate, coordinate, and verify agent actions, with machine autonomy. This balance requires allowing humans to decide when to grant or withdraw initiative, maintaining transparency into agents' reasoning processes, and preserving authority over interpretation and decision-making. After all, the goal of SciVis is not automation for its own sake, but to empower people to \hot{conduct scientific analysis, derive meaningful insights, and make informed decisions.}

Recent work in other domains highlights the promise of {\bf human-in-the-loop agentic systems}. Senoner et al.\ \cite{senoner2024explainable} showed that explainable AI enhances performance in real-world collaboration tasks, underscoring the value of interpretability. HULA~\cite{takerngsaksiri2025human} is a human-in-the-loop software framework that enables practitioners to refine plans and code iteratively, aligning automation with developer expertise. Magentic-UI~\cite{mozannar2025magentic} is an open-source platform that offers mechanisms such as co-planning, co-tasking, and action approval, emphasizing human–agent interaction over task completion.

Motivated by these insights, we propose \textbf{HiLSVA} (\underline{H}uman-in-the-\underline{L}oop \underline{S}cientific \underline{V}isualization \underline{A}gentic System), designed to bridge the gap between autonomous visualization agents and the inherently human-centered goals of SciVis. HiLSVA is grounded in the principle that automation should amplify, rather than replace, human analytical capabilities~\cite{shneiderman-maes1997debate}. By combining mixed-initiative interaction with robust agentic infrastructure, our system seeks to enhance both efficiency and \hot{transparency} in human–AI collaboration for SciVis. Specifically, HiLSVA makes the following contributions:

\textbf{Human-centered collaboration through oversight, provenance, and safe execution.}
Grounded in principles of mixed-initiative interaction~\cite{horvitz1999principle}, HiLSVA introduces fine-grained mechanisms that enable shared initiative between users and agents while preserving human authority throughout the SciVis workflow.
The system supports \emph{joint planning and execution}, where users and agents collaboratively construct explicit, stepwise plans that must be reviewed and approved before execution. 
During execution, HiLSVA treats the user as a \emph{human-in-the-loop user proxy}, enabling seamless intervention through natural language or direct manipulation of visualization tools, with fluid handoff between human and agent control. 
It further provides \emph{stepwise control and provenance tracking}, recording execution steps, software states, and visualization outputs to support undo, comparison of alternative trajectories, and reuse of validated workflows. 
To ensure reliability and safety, all agent actions are performed within \emph{sandboxed execution environments} that enforce session isolation and require explicit user approval for critical or irreversible operations. 
In addition, \emph{learn-at-test-time with human-in-the-loop feedback} enables HiLSVA to detect uncertainty, proactively query users for clarification, and incorporate feedback into a dynamic knowledge base that adapts across tasks. 
    
    

\hot{{\bf Domain-adapted integration with SciVis pipelines.}}
Rather than relying on slow, error-prone GUI-based control used by general-purpose agents~\cite{anthropic2024computeruse, zhang2025ufo2}, HiLSVA leverages MCP to integrate directly with SciVis engines via specialized, tool-aware agents. It incorporates a ChatVis-style code agent~\cite{peterka2025chatvis} for robust generation and iterative refinement of visualization scripts, as well as a ParaView agent built on ParaView-MCP~\cite{liu2025paraview} for reliable, parameterized interaction with the visualization engine. These agents provide a modular integration layer that ensures efficient, reproducible execution and allows HiLSVA to adapt to diverse visualization pipelines without altering the overall system design.

%

\textbf{Comprehensive evaluation across autonomy levels and participant expertise.}
We evaluate HiLSVA through selected case studies and a user study examining system behavior across different levels of user involvement. Twelve participants interacted with three autonomy modes in a controlled, balanced setup, ensuring that each participant experienced all modes across tasks, followed by a final task with free-mode selection. The results indicate a clear tradeoff, where higher automation improves execution efficiency, while increased human involvement enhances control and oversight.
Our study further investigates how participants with varying levels of expertise, from novices to domain scientists and SciVis experts, interact with and benefit from the system. Across all expertise levels, users were able to complete non-trivial SciVis tasks with HiLSVA support. Post-study feedback indicates strong perceived usability, transparency, and support for iterative refinement, suggesting that HiLSVA effectively supports users with diverse backgrounds, including novices.

These contributions establish HiLSVA as a practical and extensible system for human-in-the-loop SciVis. By foregrounding user agency, adaptability, and transparency, HiLSVA advances the design of SciVis agents that empower people to derive meaningful insights and \hot{scientific analysis} rather than being sidelined by automation.

%% file: src/related.tex
\section{Related Work}
\hot{
LLM agents have enabled new opportunities for SciVis through natural language interaction, workflow automation, and domain-specific analysis. Existing systems have explored autonomous SciVis pipelines, visualization-specific agents, and human-agent collaboration. As shown in Table~\ref{tab:contribution_matrix}, however, these capabilities are often studied in isolation. HiLSVA instead centers sustained human participation, combining domain-adapted SciVis agents with human oversight, mixed-initiative interaction, provenance-aware workflows, safe execution, and learn-at-test-time (LTT) adaptation from human feedback. These capabilities position agentic SciVis as a human-centered process in which autonomy supports, rather than replaces, human reasoning and exploration.}

\vspace{-0.1in}
\begin{table}[h]
\caption{\hot{Comparison of representative agentic systems.}}
\label{tab:contribution_matrix}
\vspace{-0.1in}
\centering
\resizebox{\columnwidth}{!}{
\begin{tabular}{c|cccccc}
\shortstack{method\\\ } &
\shortstack{SciVis\\adaptation} &
\shortstack{human\\oversight} &
\shortstack{mixed-\\initiative} &
\shortstack{workflow\\provenance} &
\shortstack{safe\\execution} &
\shortstack{LTT with\\human feedback}
\\ \hline
\rule{0pt}{2.4ex}
Magentic-UI~\cite{mozannar2025magentic}
& \ding{55} & \ding{51} & \ding{51} & \ding{55} & \ding{51} & \ding{55} \\

Cocoa~\cite{feng2024cocoa}
& \ding{55} & \ding{51} & \ding{51} & \ding{55} & \ding{55} & \ding{55} \\

CowPilot~\cite{huq2025cowpilot}
& \ding{55} & \ding{51} & \ding{55} & \ding{55} & \ding{55} & \ding{55} \\

VizGenie~\cite{biswas2025vizgenie}
& \ding{51} & \ding{55} & \ding{55} & \ding{55} & \ding{55} & \ding{51} \\

NLI4VolVis~\cite{ai2025nli4volvis}
& \ding{51} & \ding{55} & \ding{51} & \ding{55} & \ding{55} & \ding{55} \\

InferA~\cite{tam2025infera}
& \ding{51} & \ding{55} & \ding{55} & \ding{51} & \ding{51} & \ding{55} \\

SASAV~\cite{sun2026sasav}
& \ding{51} & \ding{55} & \ding{55} & \ding{55} & \ding{55} & \ding{55} \\

AI VIS co-scientist~\cite{miao2026toward}
& \ding{51} & \ding{51} & \ding{55} & \ding{51} & \ding{55} & \ding{55} \\

HiLSVA (ours)
& \ding{51} & \ding{51} & \ding{51} & \ding{51} & \ding{51} & \ding{51} \\

\end{tabular}
}
\vspace{-0.1in}
\end{table}

\vspace{-0.05in}
\subsection{LLM-Based Agents}

\hot{
\textbf{General-purpose agents.}
General-purpose LLM agents aim to operate across diverse environments with minimal task-specific engineering. Recent surveys identify reasoning, planning, tool use, memory, and reflection as core capabilities of such systems~\cite{xi2025rise, masterman2024landscape}. Key advances include API and external-tool use~\cite{schick2023toolformer, qin2023toolllm}, interaction with operating systems and GUIs~\cite{wu2024oscopilot, zhang2024lookchain}, search-based reasoning with backtracking~\cite{yao2023treeofthoughts, koh2024treesearch}, and self-reflection for iterative refinement~\cite{pan2024autonomous}. These capabilities allow LLMs to move beyond single-step text generation toward long-horizon, tool-grounded problem solving. More recently, self-evolving agents have explored continual adaptation beyond static tool chains. For example, SEAgent~\cite{sun2025seagent} studies autonomous software learning through curriculum-driven exploration, while Fang et al.\ \cite{fang2025comprehensive} survey broader mechanisms and challenges for lifelong agentic systems.}

\hot{
\textbf{Visualization agents.}
Visualization agents adapt general agentic capabilities to tasks where visual feedback, iterative refinement, and domain-specific evaluation are central. MatPlotAgent~\cite{yang2024matplotagent} combines query understanding, code generation, and visual feedback for scientific plotting, while LIDA~\cite{dibia2023lida} employs a grammar-agnostic pipeline for data summarization, goal exploration, and visualization specification. AVA~\cite{liu2024ava} and IntuiTF~\cite{wang2025intuitf} further demonstrate how multimodal feedback can guide low-level visualization parameter optimization. Other systems emphasize collaborative, tool-grounded, or multi-agent workflows: ChatVis~\cite{peterka2025chatvis} iteratively generates and refines ParaView scripts through dialogue, CoDA~\cite{chen2025coda} coordinates specialized agents for visualization specification and validation, and ParaView-MCP~\cite{liu2025paraview} connects LLMs to ParaView through MCP-based structured tool calls. Similar domain-specific systems have been developed for bioimage visualization and topological data analysis, such as BioImage-Agent~\cite{bioimage-agent} and TopoPilot~\cite{gorski2026topopilot}. Furthermore, SciVisAgentSkills~\cite{ai2026scivisagentskills} explores reusable agent skills for improving general-purpose coding agents across SciVis tools. Vonderhorst et al.\ \cite{vonderhorst2026exploring} compared how different agent designs and interaction modalities affect multi-step SciVis task performance. While these approaches demonstrate the promise of LLMs for visualization creation, refinement, and automation, they primarily focus on agent capability, tool integration, or output generation rather than sustained human oversight throughout SciVis workflows.}

\vspace{-0.05in}
\subsection{Human–Agent Collaboration}

Reintroducing humans into the decision loop challenges the pursuit of full automation, often introducing additional effort and coordination costs. However, empirical evidence suggests that pairing humans with agents can yield significant productivity gains compared to human-only workflows. For example, randomized controlled trials of GitHub Copilot have shown that developers complete programming tasks over 50\% faster with AI assistance~\cite{peng2023impact}. Meanwhile, studies warn that improvements in an AI system's standalone accuracy do not always translate into better team performance: model updates can break established user expectations, leading to incompatibility between human mental models and machine behavior~\cite{bansal2019updates, bansal2021does}.
Communication remains another open challenge, as humans and agents must align on intent, interpretability, and \hot{appropriate levels of human oversight} to collaborate effectively~\cite{bansal2024challenges}.

\textbf{Multi-agent collaboration.} 
Multi-agent systems provide one route for scaling coordination. Classical work emphasized robustness, modularity, and division of labor as the main benefits of distributed agents~\cite{tambe1998implementing, stone2000multiagent}. Modern approaches bring these ideas into the LLM era, enabling specialized agents to converse, plan, and solve problems collectively. Frameworks such as AutoGen~\cite{wu2023autogen} exemplify this paradigm, while recent surveys synthesize emerging methods for communication, specialization, and joint problem solving, along with open challenges in evaluation and safety~\cite{guo2024large,talebirad2023multi}.

\begin{figure*}[htb]
\centering
\includegraphics[width=0.875\linewidth]{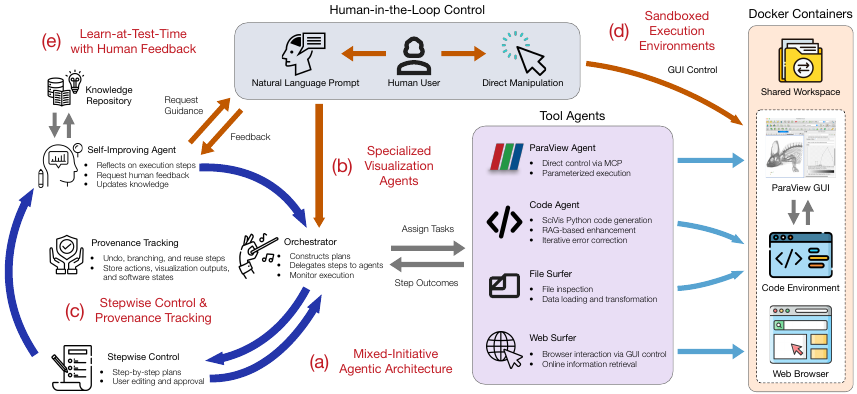}
\vspace{-0.1in}
\caption{Overview of HiLSVA.
(a) 
An orchestrator interprets user intent, constructs explicit stepwise plans, and coordinates specialized agents with bidirectional human–agent interaction.
(b) 
Specialized visualization agents execute SciVis actions via direct API calls or GUI control and report execution outcomes to the orchestrator.
(c) 
Workflows are represented as editable steps, planned and executed actions, software states, and visualization outputs, with these recorded to support undo, branching, and reuse.
(d) 
Tool interactions run in isolated Docker containers, enabling safe, parallel execution across sessions.
(e) 
A self-improving agent reflects on execution outcomes, assesses uncertainty, queries users when needed, and updates a knowledge repository during inference.}
\label{fig:overview}
\vspace{-0.1in}
\end{figure*}

\textbf{Mixed-initiative systems.} 
Beyond coordination among artificial agents, an equally important strand of research emphasizes the role of humans as integral members of the team. Horvitz's principles for mixed-initiative interfaces~\cite{horvitz1999principle} advocate balancing automation with user control, supporting uncertainty handling, dialogue, and smooth handoffs between humans and agents. Shneiderman and Maes~\cite{shneiderman-maes1997debate} emphasized that the central design tension is not between automation and control, but rather how to delegate effectively while preserving clarity, responsibility, and human judgment. Empirical work further demonstrates that interpretability enhances collaboration: explainable AI systems with visual explanations enable users to \hot{better understand and oversee model behavior}, thereby improving joint task performance compared to black-box models~\cite{senoner2024explainable}. In visualization specifically, Dhanoa et al.\ \cite{dhanoa2025agentic} argued that full automation is neither realistic nor desirable; visualization remains fundamentally human-centered, designed to amplify reasoning and support open-ended sensemaking rather than to replace human agency.

\textbf{LLM-based collaboration systems.} 
A growing line of work explores how LLM agents can share initiative with humans in open-ended tasks. Cocoa~\cite{feng2024cocoa} introduces a co-planning interface in which both the human and the agent decide on task decomposition upfront; however, the collaboration remains static, as dynamic handoffs or replanning are not supported. CowPilot~\cite{huq2025cowpilot} focuses on web navigation, providing pause, resume, and override capabilities that let humans recover from agent errors, yet lacks structured co-tasking or proactive negotiation beyond corrections. HULA~\cite{takerngsaksiri2025human} targets human-in-the-loop software development, demonstrating the benefits of pairing LLM agents with developers for code refinement, but its scope is restricted to a narrow task domain. Magentic-UI~\cite{mozannar2025magentic} advances the state of the art with fine-grained, mixed-initiative interaction: humans and agents can dynamically exchange control of subtasks, replan mid-execution, and jointly approve critical actions. 

\hot{
\textbf{Human–agent collaboration in SciVis.}
SciVis poses unique challenges for collaboration, as tasks are open-ended, visually grounded, and difficult to specify completely in advance. Recent agentic SciVis systems have explored different ways to reduce this interaction burden. VizGenie~\cite{biswas2025vizgenie} dynamically generates and validates visualization scripts to extend system functionality, while NLI4VolVis~\cite{ai2025nli4volvis} supports natural language interaction, open-vocabulary querying, and real-time editing for volume visualization. TexGS-VolVis~\cite{tang2025texgs} further enables expressive scene editing through textured Gaussian splatting. Zhang et al.\ \cite{zhang2025automatic} study semantic alignment for natural-language-driven flow visualization. InferA~\cite{tam2025infera} supports multi-agent exploration of cosmological ensemble data. More recent systems also push toward increasingly autonomous workflows: SASAV~\cite{sun2026sasav} performs self-directed scientific analysis and visualization through automated data profiling, knowledge retrieval, and parameter exploration for visualization, while AI VIS co-scientist~\cite{miao2026toward} studies an end-to-end agent harness for generating visual analysis applications from high-level task descriptions. These systems demonstrate the growing capabilities of LLM-based agents for SciVis, but primarily emphasize natural-language interaction, autonomous execution, or domain-specific analysis.}

\vspace{-0.05in}
\subsection{Evaluating Human–Agent Collaboration}  

Measuring the effectiveness of human–agent teams requires going beyond outcome metrics to capture process, interaction quality, and consistency.
\hot{
Collaborative Gym~\cite{shao2024collaborative} provides a modular environment for benchmarking human-agent teaming, while $\tau$-bench~\cite{yao2024tau} evaluates tool-augmented agents through multi-turn interactions with humans and APIs under domain-specific constraints. Fragiadakis et al.\ \cite{fragiadakis2024evaluating} further argued that evaluation should integrate both quantitative and qualitative measures to assess interpretability, transparency, and the quality of collaboration. Extending this perspective to visualization, Ai et al.\ \cite{Ai-GenAI25} emphasized systematic evaluation of agentic SciVis systems, and Do et al.~\cite{do2026svlat, Do-VISSP26} further highlighted the importance of rigorously assessing human and multimodal LLMs' understanding and interpretation capabilities in SciVis.}


Building on these insights and the human-centered nature of SciVis workflows, we evaluate HiLSVA along three axes: (1) autonomous vs.\ human-in-the-loop operation to assess mixed-initiative benefits, (2) the effect of user expertise on interaction patterns and outcomes, and (3) case studies that illustrate system behavior in realistic SciVis scenarios. \hot{Unlike fully automated benchmarks such as NL2SciVis~\cite{mathai2026nl2scivis} and SciVisAgentBench~\cite{ai2026scivisagentbench}, human-in-the-loop studies inherently involve participants with different backgrounds and expertise levels, making large-scale reproducible benchmark-style evaluation difficult. Therefore, we focus on controlled user studies and qualitative analyses that capture the collaborative nature of human–agent SciVis workflows.}

%% file: src/method.tex
\section{Our Approach}

\begin{figure*}[htb]
\centering
\includegraphics[width=0.975\linewidth]{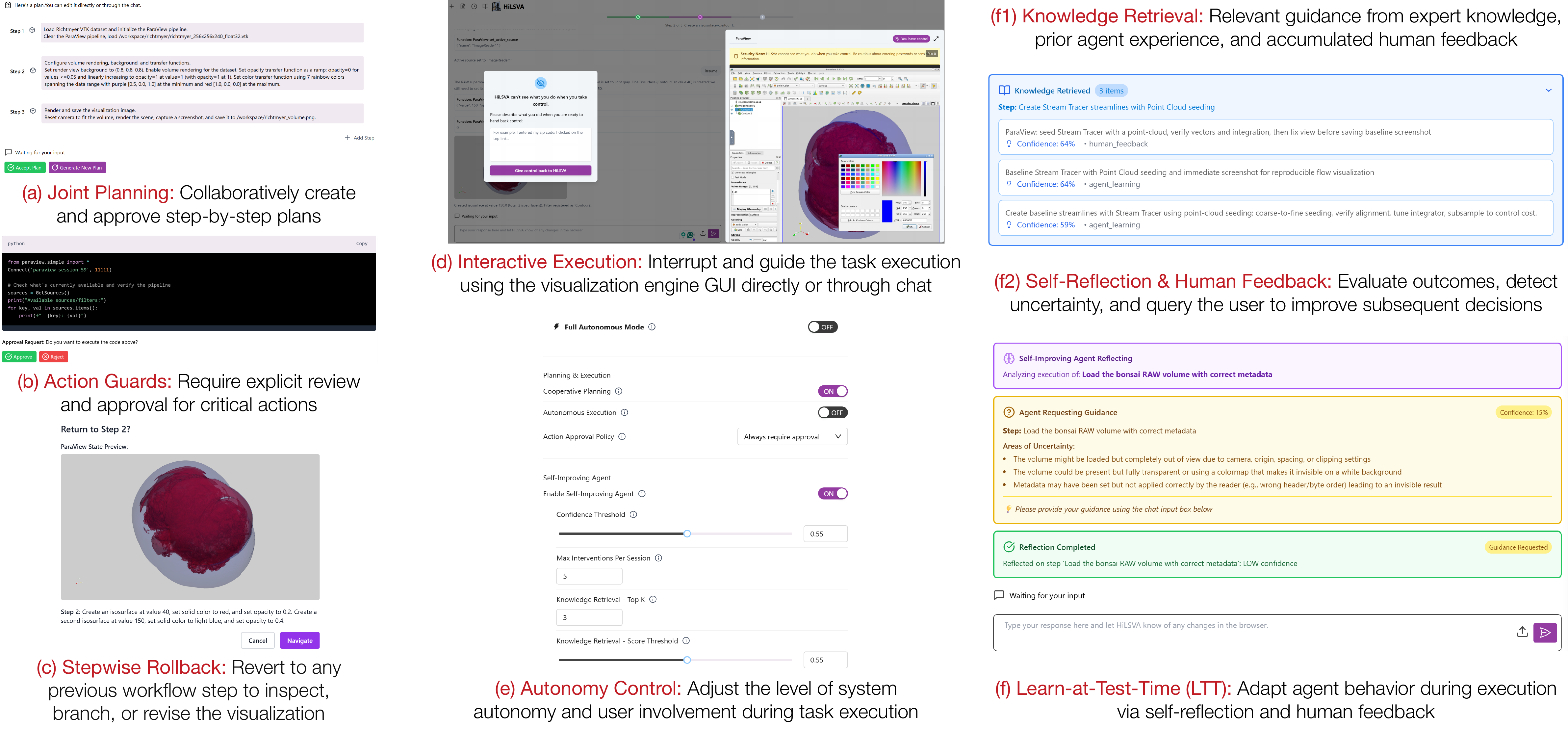}
\vspace{-0.1in}
\caption{
Key mixed-initiative capabilities of HiLSVA. The system supports (a) collaborative planning, (b) guarded execution, (c) stepwise rollback, (d) interactive user steering, (e) autonomy control, and (f) LTT adaptation through retrieval, reflection, and human feedback.
}
\label{fig:example_features}
\vspace{-0.1in}
\end{figure*}

\hot{We implement HiLSVA on top of Magentic-UI~\cite{mozannar2025magentic}, which provides the general mixed-initiative infrastructure used for planning, execution, and action approval. As illustrated in Figure~\ref{fig:overview}, HiLSVA extends this foundation with SciVis-specific capabilities, including specific agent designs, provenance-aware execution with human oversight, multimodal visualization control via MCP tool calls, generated scripts, and direct user interaction, sandboxed environments, and test-time adaptation via human feedback.} 
This section details these design choices across mixed-initiative agentic architecture, mechanisms for human oversight and safe execution, LTT adaptation, and interactive interface design. \hot{Although the current implementation centers on ParaView, the architecture is designed to support multiple SciVis backends.} In Figure~\ref{fig:example_features}, we highlight the key mixed-initiative capabilities of HiLSVA.

\vspace{-0.05in}
\subsection{Mixed-Initiative Agentic Architecture}

{\bf Multi-agentic system design and orchestration.}
HiLSVA adopts a multi-agent architecture centered around a lead \textbf{orchestrator} that coordinates a set of specialized sub-agents. These include a \textbf{web surfer} for browser-based information seeking, a \textbf{code agent} for generating and executing Python and shell scripts to manipulate SciVis tools, a \textbf{file surfer} for file inspection and transformation, a \textbf{ParaView agent} that directly controls the ParaView GUI via an MCP server, and a \textbf{self-improving agent} that supports test-time learning through human feedback. \hot{These agents were selected to support backend-agnostic SciVis workflows. The orchestrator interprets user intent, constructs an explicit plan, assigns each step to an appropriate agent, tracks execution progress, and generates the final response to the user. The ParaView agent and code agent provide standard interfaces to visualization software, enabling the architecture to be extended to different backends, while the remaining agents extend the system with external information access and experience-driven adaptation.}

The system follows a plan-first workflow. Upon receiving a task, the orchestrator synthesizes a stepwise plan, with each step assigned to a specific agent. Rather than relying on multi-turn dialogue for refinement, HiLSVA exposes the plan as an editable artifact. Users may reorder steps, add or remove actions, or modify step instructions before execution, enabling precise and low-effort control over the planning stage (Figure~\ref{fig:example_features}(a)). Execution begins only after the user approves the plan.
Once execution starts, HiLSVA operates in a mixed-initiative manner. At each step, the orchestrator monitors execution status, determines whether the step has been completed, and may revise instructions or trigger replanning when necessary. Before executing a step, the orchestrator may invoke the self-improving agent to retrieve relevant domain knowledge from an external knowledge base. After each step, the self-improving agent can reflect on outcomes and, when appropriate, solicit additional human guidance to refine subsequent actions. 

{\bf Specialized visualization agents.}
In our current implementation, HiLSVA runs ParaView~\cite{Ahrens2005ParaView} inside isolated Docker environments, ensuring reproducible and safe execution. \hot{We selected ParaView because it is one of the most comprehensive and widely adopted SciVis platforms, providing a rich set of visualization and analysis capabilities while uniquely supporting multiple interaction modalities on the same visualization instance, including MCP tool calls, generated Python scripts, and direct GUI manipulation by the user, enabling seamless handoff between human and agent control.}

The \emph{code agent} in HiLSVA adopts techniques similar to ChatVis~\cite{peterka2025chatvis}, including structured prompt decomposition, retrieval-augmented generation (RAG) over documentation and code examples, and iterative error checking with feedback-driven refinement until executable visualization scripts are produced. This design enables robust generation and correction of SciVis code scripts that involve complex tool APIs.
To support reliable tool interaction, the \emph{ParaView agent} interfaces with the visualization engine via ParaView-MCP~\cite{liu2025paraview}. The MCP server provides a higher-level, function-oriented abstraction over ParaView’s Python API, enabling the agent to invoke visualization operations via parameterized function calls rather than low-level scripting.

\hot{The architecture itself is backend-agnostic. Visualization tools are encapsulated within sandboxed containers and accessed through specialized agents. Consequently, other domain-specific visualization environments can be integrated without changing the overall system design. Examples include VMD~\cite{VMD1996} for molecular visualization, napari~\cite{napari2019} for bioimage visualization, and TTK~\cite{tierny2017topology} for topology analysis. If corresponding MCP servers or tool interfaces (e.g., GMX-VMD-MCP~\cite{egtai2025gmxvmdmcp}, BioImage-Agent~\cite{bioimage-agent}, TopoPilot~\cite{gorski2026topopilot}) are available, they can be incorporated as additional specialized agents within the existing framework.}

\vspace{-0.05in}
\subsection{Human Oversight, Provenance, and Safe Execution}

{\bf Human-in-the-loop user proxy.}
The user is treated as a first-class participant in the multi-agent team, and the orchestrator may delegate specific steps to the user as part of the shared execution process. During execution, HiLSVA supports mixed-initiative co-execution~\cite{mozannar2025magentic, feng2024cocoa, huq2025cowpilot}. Users may intervene at any time through natural-language instructions or direct interaction with the visualization environment, such as adjusting transfer functions or camera parameters when visual intent is difficult to specify in words. After intervention, users can return control to the agent, enabling fluid handoff between manual and automated execution (Figure~\ref{fig:example_features}(d)).
Initiative also flows in the opposite direction: when the orchestrator or a specialized agent detects ambiguity, deviation from the approved plan, or insufficient confidence to proceed, the system proactively queries the user for clarification. Moreover, HiLSVA enforces explicit user approval for critical or potentially irreversible actions, such as executing generated code or sensitive SciVis tool calls (Figure~\ref{fig:example_features}(b)). Users can also adjust the level of system autonomy, balancing agent autonomy and human intervention (Figure~\ref{fig:example_features}(e)).

\textbf{Stepwise control and provenance tracking.}
Reproducibility and accountability are fundamental requirements in scientific research and are closely tied to provenance~\cite{freire2008provenancesurvey}. Provenance provides a structured record of how results are produced, enabling interpretation, verification, and reuse of computational workflows. Prior work in visualization and data analysis has demonstrated the value of provenance for supporting sensemaking, reflection, and responsible system behavior~\cite{bavoil2005vis, werder2022tmis, narechania2025tvcg}.

HiLSVA provides fine-grained, stepwise control through an explicit workflow representation, allowing users to revisit and revise their analysis as it unfolds. For each execution step, the system records both {\em prospective} provenance (the planned
sequence of actions) and {\em retrospective} provenance (the executed actions, software states, and visualization outputs). Users can select any prior step to restore the corresponding state, modify the remaining plan from there, and continue execution along the updated workflow plan (Figure~\ref{fig:example_features}(c)).
In addition to tracking individual executions, HiLSVA supports storing and reusing complete workflow plans. After a task is completed, users may choose to save the validated plan along with its associated execution provenance. Saved plans can be replayed on the original task, used as guidance for new tasks, or automatically retrieved by the orchestrator.

\hot{Unlike plan-level retrieval, this workflow reuse is stateful. Users can restore an arbitrary step within a saved workflow together with the corresponding visualization software state, inspect intermediate results, branch from that point, and continue execution with revised instructions. This enables provenance-aware reuse and refinement of entire SciVis workflows rather than only reusing planning artifacts.}


\textbf{Sandboxed execution environments.}
HiLSVA employs plug-and-play sandboxed execution to ensure safety and reproducibility. Each dialogue session runs within its own isolated Docker environment for the visualization engine (i.e., ParaView), code execution, web browsing, and file access. These containers share a controlled workspace mapped to the local file system, enabling coordinated interaction among agents while preventing unintended side effects across sessions.
HiLSVA also supports parallel execution of multiple tasks, allowing users to safely experiment with different settings without compromising the host system. Users can switch between active sessions, each backed by an independent ParaView Docker instance, with a session status indicator that indicates whether user input is required.

\vspace{-0.05in}
\subsection{LTT with Human-in-the-Loop Feedback}
\label{subsec:LTT}

LTT refers to a model's ability to adapt its behavior during inference after deployment. For LLMs, common LTT mechanisms include in-context or few-shot learning~\cite{brown2020nips, dong2024emnlp}, RAG~\cite{lewis2020nips}, and, more recently, test-time fine-tuning~\cite{hubotter2024efficiently}. In the context of agentic systems, self-learning agents aim to improve autonomously through interaction with their environment~\cite{sun2025seagent, fang2025comprehensive}. HiLSVA builds on this line of work by explicitly incorporating humans into the test-time learning loop~\cite{he2025enabling}, enabling agents to adapt through structured self-reflection and targeted human feedback during task execution (Figure~\ref{fig:example_features}(f)).

{\bf Problem formulation.}
HiLSVA follows a plan-first-then-execute workflow. For a given task, the agent team executes a sequence of steps
$\mathcal{S} = (s_1, s_2, \ldots, s_N)$,
where each step corresponds to a dynamically evolving environment
$\mathcal{E} = (e_1, e_2, \ldots, e_N)$.
To complete step $s_i$, the agents perform a sequence of actions $a_i \in \mathcal{A}$. The environment state $e_i$ includes the visualization engine state, auxiliary tool states, intermediate files, and relevant knowledge retrieved from an external repository. To operate effectively in this changing environment, the agent team maintains an internal state $\Theta$ and receives guidance from a human user $\mathcal{H}$ within a predefined interaction budget $B$.

{\bf LTT.}
During execution of step $s_i$, the agent team maintains an internal state $\theta_i \in \Theta$ and follows a decision policy
$\pi : \mathcal{E} \times \Theta \rightarrow \mathcal{A}$,
which produces actions $a_i = \pi(e_i, \theta_i)$. We model LTT as an update of the internal state after each step
$\theta_{i+1} = f(\theta_i, e_i, a_i, r_i; q_i, h_i)$,
where $r_i$ denotes the agent’s self-reflection after completing step $s_i$, $q_i$ is a clarification query posed to the user, and $h_i$ is the corresponding human feedback.

{\bf Knowledge updating with human feedback.}
In HiLSVA, the internal state of the agent team is instantiated as a vector-based knowledge repository $R_i$, i.e., $\theta_i \approx R_i$. Each knowledge item is represented by a natural language caption and indexed using a BERT-based embedding of that caption. The initial repository $R_1$ is seeded with general SciVis knowledge, and is incrementally augmented with insights derived from prior task executions and human feedback.

Before executing step $s_i$, relevant knowledge items are retrieved from $R_i$ and incorporated into the environment state $e_i$ to guide agent actions. Retrieval is based on a composite score defined as
$\text{score} = w_{\text{valid}} \cdot s_{\text{semantic}} \cdot  s_{\text{confidence}} \cdot s_{\text{recency}}$,
where $w_{\text{valid}} \in \{0,1\}$ denotes a validity weight indicating whether an item is considered outdated, $s_{\text{semantic}}$ is the cosine similarity between the BERT embedding of the step description for $s_i$ and the embedding of the knowledge caption, and $s_{\text{confidence}} \in [0,1]$ reflects the reliability of the knowledge item.
Knowledge derived from human feedback is assigned a confidence score of $1$, while insights obtained through self-reflection receive confidence scores graded by the self-improving agent based on its assessment of action reliability and execution outcomes.
The recency score is defined as an exponential decay,
$s_{\text{recency}} = \exp(-\lambda \cdot \Delta t)$,
where $\Delta t$ is the normalized time since the item's last validation and $\lambda$ is a decay rate. Certain foundational knowledge items (e.g., core SciVis principles) are marked as non-expiring and assigned a recency score of $1$.

After executing step $s_i$, the self-improving agent performs a self-reflection to assess execution quality, assigns a confidence score, and summarizes any ambiguities requiring clarification as a query $q_i$. 
Each query incurs a cost $c(q_i) > 0$, set uniformly to $1$ in our experiments for simplicity. In principle, different query types could be assigned different costs depending on their cognitive or interaction burden.
If the confidence score falls below a predefined threshold and the accumulated cost satisfies $\sum_{j=1}^{i} c(q_j) \le B$, the agent requests human feedback $h_i = \mathcal{H}(e_i, a_i, q_i)$.
The knowledge repository is then updated according to
$R_{i+1} = g(R_i, r_i, q_i, h_i)$,
where the update function $g$ constitutes the core of the LTT function $f$. $g$ stores both self-reflection summaries and human feedback as new knowledge items, along with associated metadata including timestamps, confidence scores, and LLM-generated captions. These metadata are subsequently used to support informed retrieval in future steps.

{\bf Self-improving agent and user expertise.}
\hot{LTT with human feedback in HiLSVA is implemented via retrieval-based adaptation rather than model weight updates or architectural changes. A dedicated self-improving agent observes all actions and environment states, performs confidence assessment and self-reflection after each execution step, and decides whether to request human input. Validated insights and self-reflections are stored as natural-language knowledge items associated with confidence and recency scores. These knowledge items are retrieved to enable adaptation through accumulated experience and human feedback.}
This mechanism can be disabled to support ablation studies that quantify its impact on task performance. HiLSVA is designed to collaborate with users of varying expertise levels, including non-experts, domain experts, and SciVis experts; accordingly, our evaluation examines how user expertise influences interaction patterns, collaboration dynamics, and outcome quality.

\begin{figure}[htb]
\centering
\includegraphics[width=1.0\linewidth]{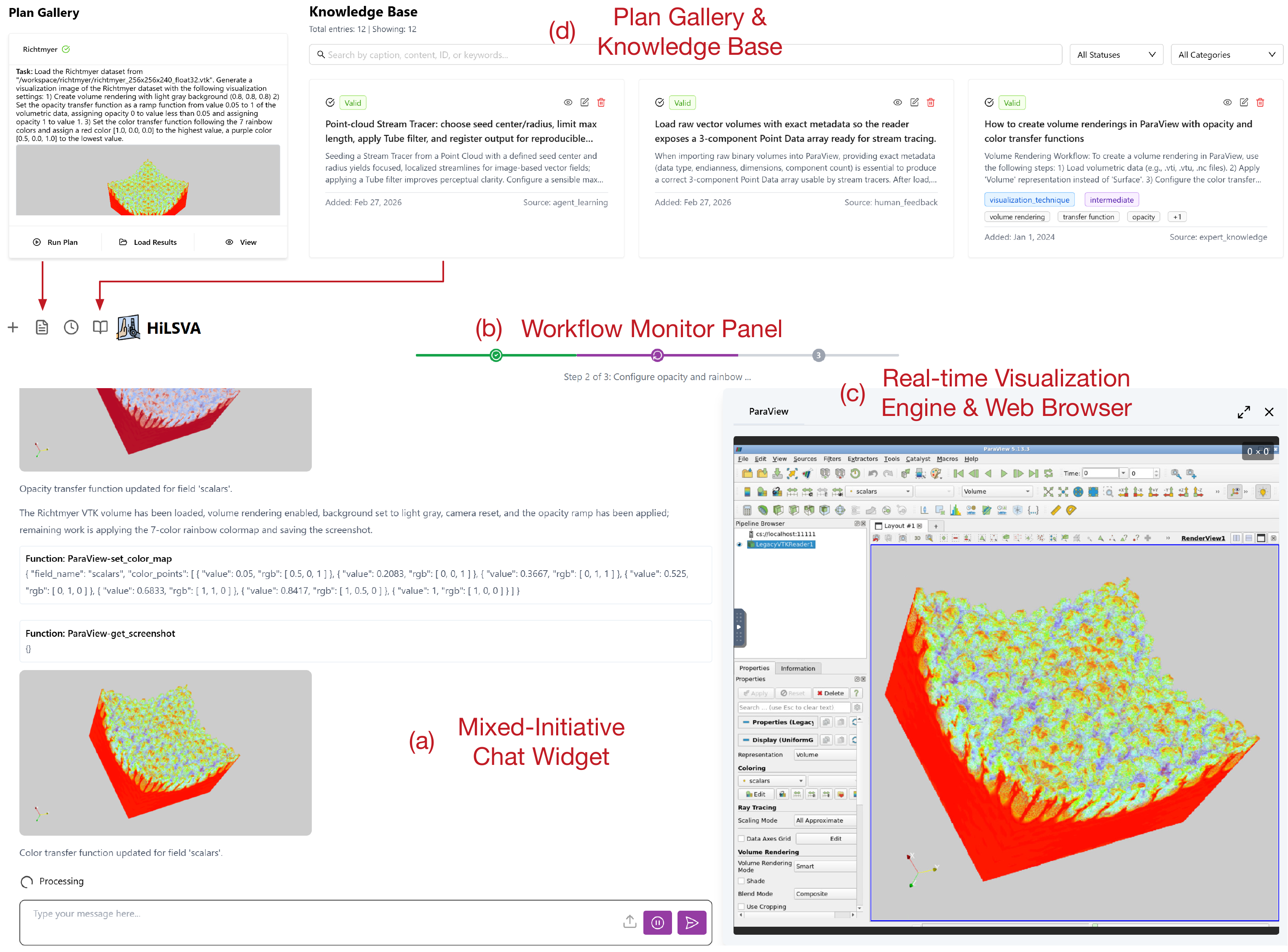}
\vspace{-0.2in}
\caption{
Overview of the HiLSVA interface. 
(a) \emph{Mixed-initiative chat widget} for natural language planning, step refinement, and uncertainty prompts. 
(b) \emph{Workflow monitor panel} that displays execution progress and step status, and supports clickable state transitions for returning to prior steps.
(c) \emph{Real-time visualization engine} and \emph{web browser}, where the agent's actions and user interventions directly operate on the ParaView rendering environment. 
(d) \emph{Plan gallery} and \emph{knowledge base} that accumulate validated workflows and learned experience to support provenance-aware reuse and LTT adaptation.
}
\label{fig:GUI}
\vspace{-0.1in}
\end{figure}

%

\vspace{-0.05in}
\subsection{Interactive Interface}

HiLSVA provides an interface that tightly couples planning, execution, and visualization. The interface consists of three primary components: a {\em mixed-initiative chat widget} for natural-language interaction, a {\em workflow monitor panel} for tracking execution progress and provenance, and a {\em real-time visualization engine} and a {\em web browser}. As shown in Figure~\ref{fig:GUI}, these components operate in synchrony to support collaborative SciVis workflows.
The mixed-initiative chat widget serves as the primary interaction channel for user–agent interaction. Through chat, users can specify tasks, inspect and edit execution plans, approve or reject actions, review generated code, and provide feedback in response to agent self-reflection.
The workflow monitor panel exposes the current execution state as a stepwise process. Users can inspect progress, undo completed steps, and return to prior states using the recorded provenance.
In parallel, the visualization engine and the web browser update in real time as agents interact with external tools. Users may intervene directly by taking control of the GUI at any point, then seamlessly resume automated execution.
%
In addition to these core components, HiLSVA includes a {\em plan gallery} and a {\em knowledge base} that present previously saved workflows, accumulated knowledge, and their associated provenance, as well as a {\em session monitor sidebar} that allows users to manage multiple concurrent task sessions and view their execution status.

%% file: src/results.tex
\section{Results and Evaluation}

This section presents our approach to analyzing and evaluating HiLSVA. We combine structured case studies, a controlled user study followed by surveys and interviews, and multimodal analysis integrating benchmark-driven quantitative metrics and qualitative evaluations. Together, these methods enable a comprehensive investigation of how mixed-initiative agentic systems can support data exploration, visualization construction, and \hot{scientific analysis}.
For the backbone LLM, we adopt GPT-5.2 for the orchestrator and Claude-Sonnet-4.6 for the other tool agents.

The Institutional Review Board at our University approved the user study conducted in this work.
\hot{Before accessing the study materials, all participants reviewed and signed an informed consent form.} They received an explanation of the study's objectives and procedures, and consent was documented by completing a hard-copy consent form.

\vspace{-0.05in}
\subsection{SciVis Stages and Case Study Design}

Our evaluation is structured around three canonical SciVis stages: {\em data exploration}, {\em data visualization}, and {\em scientific insight}, which guide us in designing case studies of increasing analytic complexity.
In the {\bf data exploration} stage, users familiarize themselves with unfamiliar multidimensional datasets by selecting visualization techniques, identifying salient structures, and determining how to begin analysis. This aligns with our {\bf basic action tasks}, which test HiLSVA's reliability in executing foundational operations such as loading data, adjusting views, filtering, and generating renderings.
The {\bf data visualization} stage focuses on developing iterative, mixed-initiative workflows to address analytical questions. Here, both user and agent collaborate through planning, refinement, and clarification. This corresponds to our {\bf visualization workflow tasks}, which require multi-step reasoning, coordinated pipeline construction, and negotiation to isolate regions of interest, refine transfer functions, or generate comparative views.
Finally, the {\bf scientific insight} stage focuses on moving beyond direct visualization of raw data to derive, compare, and interpret scientifically meaningful features. This parallels our {\bf \hot{scientific analysis} tasks}, which involve analysis-driven, open-ended objectives such as computing derived quantities, identifying structural patterns, and extending workflows across different simulation conditions. HiLSVA's adaptive behavior and the ability to reuse previously learned workflows are particularly critical at this stage.
By aligning evaluation stages with progressively complex task classes, we systematically examine when and how human–agent collaboration enhances analytic reasoning.

\vspace{-0.05in}
\subsection{Case Studies}


We present five case studies of varying complexities to showcase HiLSVA's capabilities. These include two basic action tasks (foot and hurricane), two visualization workflow tasks (tornado and combustion), and one \hot{scientific analysis} task (half-cylinder). 
\hot{For brevity, the case studies begin after a visualization strategy has been selected. In practice, HiLSVA supports co-planning to help users identify suitable visualization and analysis techniques from high-level goals.}

\begin{figure}[htb]
\centering
\includegraphics[width=1.0\linewidth]{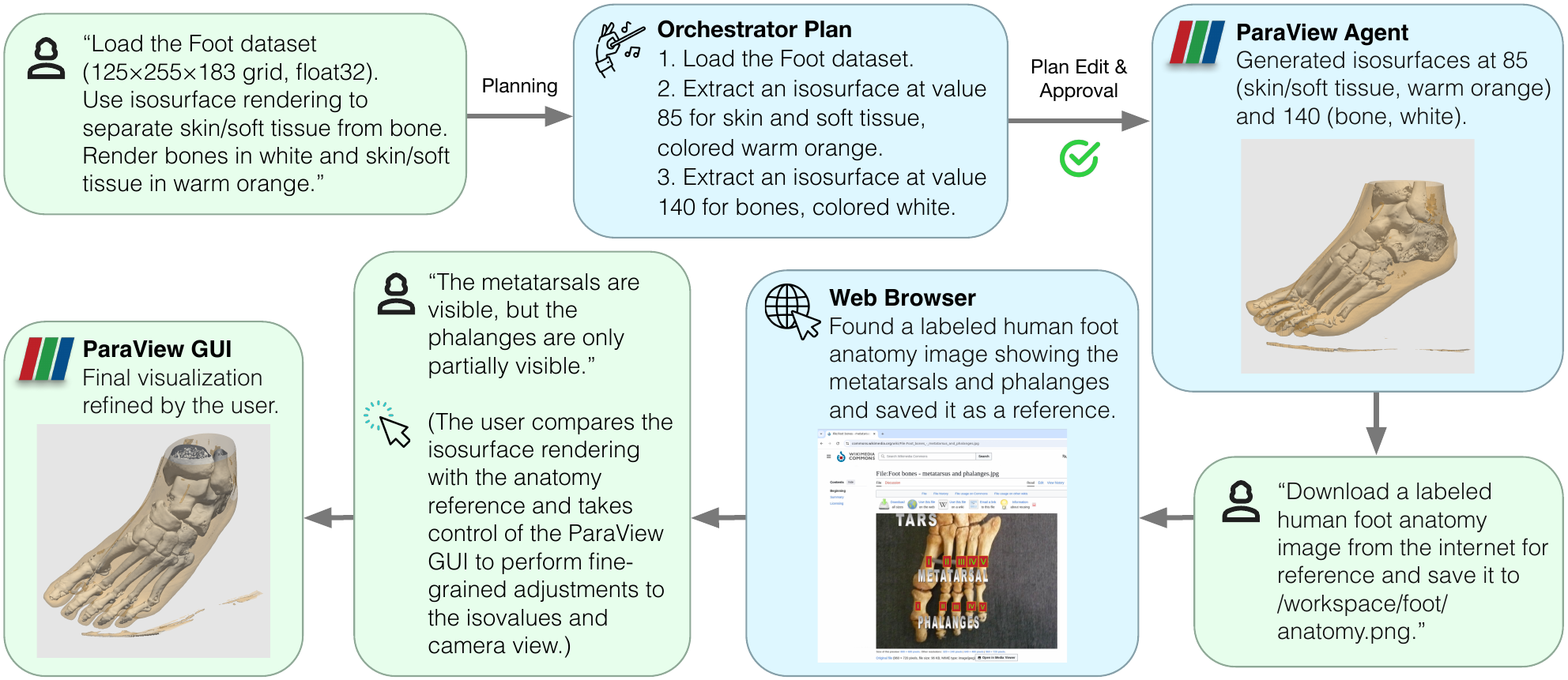}
\vspace{-0.2in}
\caption{
Case study of the foot (basic action task).
HiLSVA generates an initial isosurface visualization, retrieves an anatomy reference from the web, and supports user-driven refinement through direct GUI interaction.
}
\label{fig:foot}
\vspace{-0.1in}
\end{figure}

{\bf Foot.} 
The dataset is a CT scan of a human foot with a resolution of 125$\times$255$\times$183. 
We use this example to illustrate a basic action task. 
As shown in Figure~\ref{fig:foot}, the user first requests an initial isosurface visualization to reveal bone structures while distinguishing bone from surrounding skin and soft tissue. 
The HiLSVA orchestrator generates a plan that loads the dataset, extracts isosurfaces that separate bone from surrounding tissue, and assigns colors to highlight anatomical structures. 
After the plan is approved, the ParaView agent executes these steps and produces an initial isosurface visualization.

Next, the user asks HiLSVA to download a labeled anatomy image of the human foot showing the metatarsals and phalanges. 
The web surfer retrieves this reference from the internet, providing anatomical context that is not directly available from the raw visualization. 
By comparing the current isosurface with the anatomical reference, the user observes that the metatarsals are clearly visible, whereas the phalanges are only partially visible. 
Further refinement through natural language alone would require specifying detailed parameters, such as precise isovalues or camera adjustments, which can be cumbersome. 
Instead, HiLSVA's mixed-initiative interaction allows the user to directly control the ParaView GUI and make fine-grained adjustments to the visualization. 
Through this interactive refinement, the user quickly improves the isosurface rendering and makes the anatomical structures more clearly visible.
\hot{This example demonstrates how HiLSVA supports data exploration through iterative visualization refinement and contextual information retrieval.}

\begin{figure}[htb]
\centering
\includegraphics[width=1.0\linewidth]{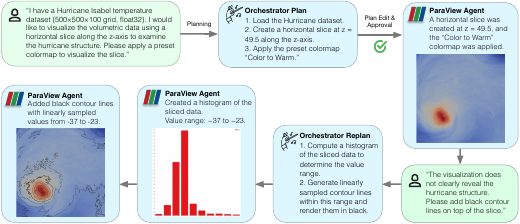}
\vspace{-0.2in}
\caption{
Case study of the hurricane (basic action task).
HiLSVA first constructs a horizontal slice of the temperature field, then analyzes the data distribution and overlays contour lines to better reveal the temperature distribution.
}
\label{fig:isabel}
\vspace{-0.1in}
\end{figure}

{\bf Hurricane.}
The dataset is an atmospheric simulation of Hurricane Isabel with a resolution of 500$\times$500$\times$100. 
In this example, we use the temperature field at a single time step to examine the hurricane structure. 
We use this example to demonstrate a basic action task supported by HiLSVA. 
As shown in Figure~\ref{fig:isabel}, the user first requests a visualization of the hurricane by creating a horizontal slice along the $z$-axis and applying a preset colormap. 
The HiLSVA orchestrator generates a plan that loads the dataset, extracts a horizontal slice at $z=49.5$, and applies the preset \textsf{Color to Warm} colormap. 
After plan approval, the ParaView agent executes these steps and produces an initial slice visualization.

However, the resulting image does not clearly reveal the hurricane structure.
The user therefore asks the system to overlay contour lines on the slice to highlight the temperature distribution better.  
In response, the orchestrator first computes a histogram of the sliced data to determine the value range, then generates contour lines using linearly sampled values within this range. 
The ParaView agent executes these steps and overlays black contour lines on the slice, making the hurricane's structure and temperature gradients more clearly visible.
\hot{This example illustrates how HiLSVA supports exploratory analysis by refining visual representations to reveal salient temperature structures.}

\begin{figure}[htb]
\centering
\includegraphics[width=1.0\linewidth]{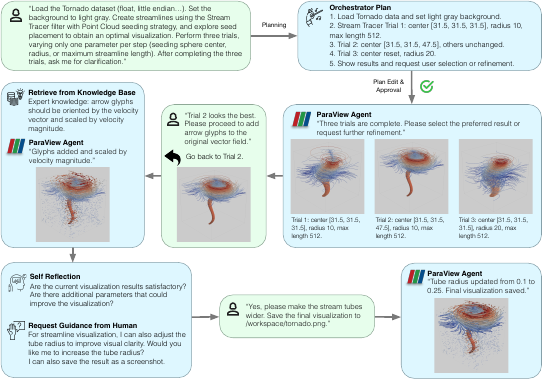}
\vspace{-0.2in}
\caption{
Case study of the tornado (visualization workflow task).
HiLSVA explores parameters for visualizing streamlines, supports rolling back to the selected trial, and adds knowledge-guided glyph visualization. 
}
\label{fig:tornado}
\vspace{-0.1in}
\end{figure}

{\bf Tornado.}
The dataset is a vector field representing a simulated tornado, with a resolution of 64$\times$64$\times$64. We use it to show a visualization workflow task supported by HiLSVA. As shown in Figure~\ref{fig:tornado}, the user first specifies basic requirements and asks the system to load the dataset. The user then requests a streamline visualization using ParaView's \textsf{Stream Tracer} filter with a \textsf{Point Cloud} seeding strategy. However, the user is uncertain about the optimal parameter configuration, including the seeding sphere center, radius, and maximum streamline length, and therefore wishes to explore alternative settings.

In response, the HiLSVA orchestrator generates a five-step plan. The workflow begins by loading the dataset and setting the background color. The system then performs three parameter trials. Trial 1 places the seeding sphere at the volume's center [31.5, 31.5, 31.5] (the data extent is [0, 63]) with radius 10 and maximum streamline length 512. Trial 2 shifts the seeding sphere's center to [31.5, 31.5, 47.5] while keeping other parameters fixed. Trial 3 moves the seeding sphere's center back to the volume's center and increases its radius to 20. HiLSVA presents the resulting visualizations and prompts the user to select the preferred configuration or request further refinement.

After inspecting the results, the user identifies Trial 2 as optimal and requires no further tuning. Using HiLSVA's provenance-aware workflow monitor, the user can instantly restore the ParaView state corresponding to Trial 2 with a single click. All parameters and rendering states are faithfully recovered. Building on the selected streamline visualization, the user next requests a \textsf{Glyph} filter on the original vector field using arrow glyphs. The self-improving agent consults the knowledge repository and retrieves expert guidance indicating that glyphs should be oriented by the velocity vector and scaled by velocity magnitude. Guided by this knowledge, the orchestrator generates a new plan and instructs the ParaView agent to produce colored arrow glyphs scaled by velocity magnitude using the same color map. 

After generating the glyph visualization, the self-improving agent performs a self-reflection step to evaluate the results and identify parameters that could further improve clarity. It suggests adjusting the streamline tube radius and asks the user for confirmation. The user agrees and requests wider stream tubes. The ParaView agent then applies a \textsf{Tube} filter and increases the tube radius from 0.1 to 0.25, and saves the final visualization.

Finally, the user may optionally preserve the workflow for future reuse. HiLSVA stores not only the stepwise plan but also the full provenance record, including ParaView states and intermediate visualization snapshots. These artifacts can be reloaded in later sessions or referred to in other sessions. 
\hot{This case demonstrates how mixed-initiative interaction helps users evaluate alternatives and refine visualization designs.}

\begin{figure}[htb]
\centering
\includegraphics[width=1.0\linewidth]{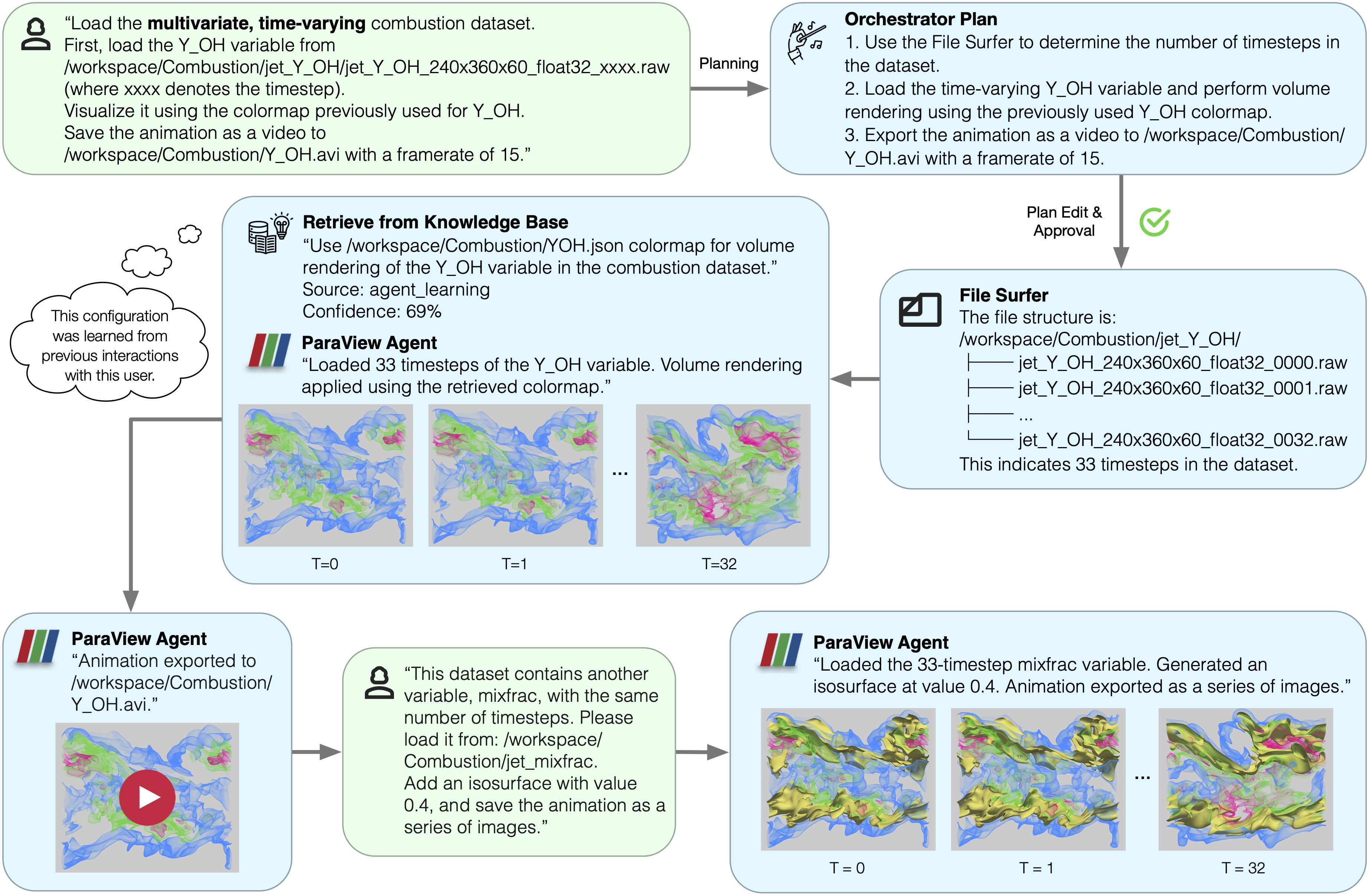}
\vspace{-0.2in}
\caption{
Case study of the combustion (visualization workflow task).
HiLSVA visualizes the time-varying \texttt{Y\_OH} scalar field using volume rendering, exports an animation, and then loads the \texttt{mixfrac} scalar field to generate an isosurface visualization across all timesteps.
}
\label{fig:combustion}
\vspace{-0.1in}
\end{figure}

{\bf Combustion.}
The dataset is a multivariate, time-varying simulation of turbulent reacting flow. In this example, we use two scalar variables: \texttt{mixfrac}, representing the mixture fraction of fuel and oxidizer, and \texttt{Y\_OH}, the mass fraction of the hydroxyl radical, a commonly used indicator of reaction zones in combustion. We use a downsampled version of the dataset containing 33 timesteps, with a resolution of 240$\times$360$\times$60.
As shown in Figure~\ref{fig:combustion}, the user first requests visualization of the time-varying \texttt{Y\_OH} variable and asks the system to export the animation as a video. The HiLSVA orchestrator generates a plan comprising three steps: determining the number of timesteps in the dataset, loading the time-varying \texttt{Y\_OH} field, and performing volume rendering using the previously used colormap before exporting the animation. To determine the number of timesteps, the file surfer inspects the directory structure and identifies 33 raw data files corresponding to individual timesteps.

Before visualization, the self-improving agent retrieves a previously learned configuration from the knowledge repository, suggesting the use of a dedicated colormap for volume rendering. This configuration was learned from prior interactions with the user and stored as reusable knowledge, enabling HiLSVA to adapt visualization parameters to user-specific preferences. Guided by this retrieved knowledge, the ParaView agent loads all timesteps of the \texttt{Y\_OH} field and performs volume rendering across the sequence. The resulting frames are then exported as a video at the requested frame rate.

After the animation is generated, the user requests further analysis using another variable in the dataset, \texttt{mixfrac}, which shares the same temporal resolution. The user asks the system to load the time-varying field and extract an isosurface with an isovalue of 0.4. The ParaView agent loads the 33 \texttt{mixfrac} timesteps, generates an isosurface visualization for each, and exports the animation as an image sequence.
\hot{This example highlights how HiLSVA supports iterative workflow refinement for multivariate, time-varying datasets.}

{\bf Half-cylinder.} 
The dataset is a small ensemble of numerical simulations of incompressible 3D flow around a half-cylinder. Each ensemble member corresponds to a different Reynolds number, and the simulations were resampled from adaptive unstructured grids onto a regular grid. In this example, we use a single timestep from the Reynolds number $Re=640$ case on a regular grid of 640$\times$240$\times$80. We use this example to demonstrate a \hot{scientific analysis} task supported by HiLSVA. \hot{Unlike the previous workflow-oriented examples, this case focuses on analysis-driven exploration. Users are not required to know specific vortex-identification techniques, such as the Q-criterion or
$\lambda_2$ in advance. Instead, users may begin with high-level goals, such as exploring flow structures or identifying regions of complex behavior. Through co-planning and iterative dialogue, HiLSVA can propose candidate analysis strategies, explain their purpose, and execute the corresponding workflow. Domain experts may choose to specify particular analysis methods directly, which can further accelerate the process, but such knowledge is not required to initiate the analysis.} 

\begin{figure}[htb]
\centering
\includegraphics[width=1.0\linewidth]{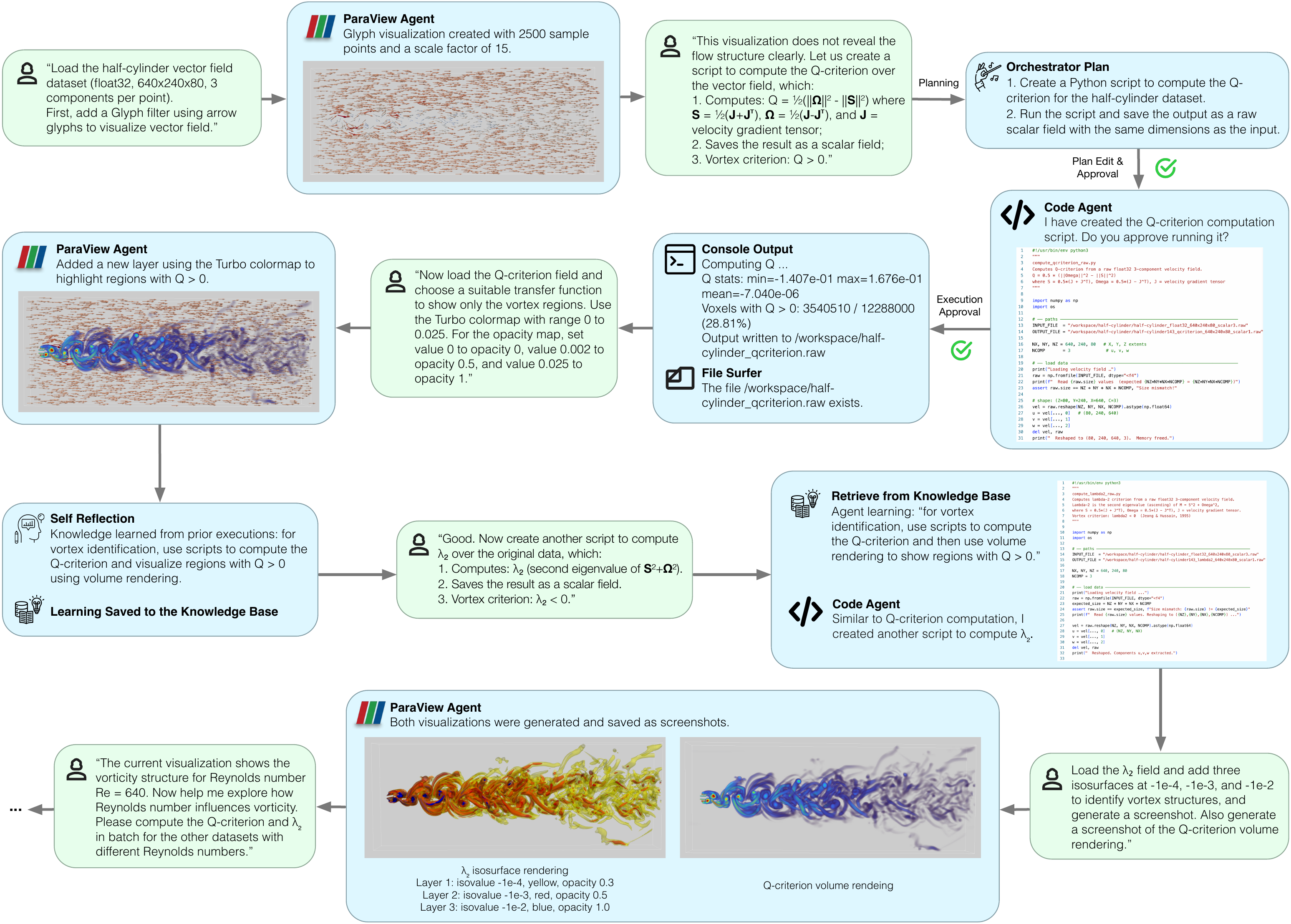}
\vspace{-0.2in}
\caption{
Case study of the half-cylinder (\hot{scientific analysis} task).
HiLSVA begins with a glyph-based visualization of the vector field, then computes and visualizes the Q-criterion and $\lambda_2$ fields to reveal vortex structures. The system further saves this successful workflow as reusable knowledge and supports extending the analysis to additional Reynolds numbers.
}
\label{fig:half-cylinder}
\vspace{-0.1in}
\end{figure}

As shown in Figure~\ref{fig:half-cylinder}, the user first requests a glyph-based visualization of the vector field using arrow glyphs. The ParaView agent then loads the dataset and generates the initial glyph visualization. However, the resulting visualization does not clearly reveal the underlying flow structures. Rather than stopping at this initial rendering, the user asks the system to help identify vortex regions through more scientifically meaningful analysis. This shift from direct depiction to derived-field analysis exemplifies a characteristic of \hot{scientific analysis} tasks: the goal is not only to visualize the data but also to uncover latent structures that support interpretation. In response, HiLSVA generates a plan to compute the Q-criterion, asks for approval before executing the generated script, and saves the resulting scalar field. The script computes
$Q=\tfrac{1}{2}\left(\|\bm{\Omega}\|^2-\|\mathbf{S}\|^2\right)$,
where $\mathbf{S}=\tfrac{1}{2}(\mathbf{J}+\mathbf{J}^\top)$ and $\bm{\Omega}=\tfrac{1}{2}(\mathbf{J}-\mathbf{J}^\top)$ are the symmetric and antisymmetric parts of the velocity gradient tensor $\mathbf{J}$. The ParaView agent then loads the computed Q-criterion field and applies volume rendering with a user-specified transfer function to highlight regions with $Q>0$, revealing coherent vortex structures much more clearly than the initial glyph rendering.

Next, the user asks HiLSVA to analyze the flow further using the $\lambda_2$ method, another vortex identification measure. Similar to the previous step, the system generates and executes a second script to compute $\lambda_2$ from the original vector field, saves the result as a scalar field, and loads it into ParaView. The ParaView agent then creates three isosurfaces at different negative isovalues to visualize vortex structures at multiple strengths. Compared with the Q-criterion volume rendering, the $\lambda_2$ visualization offers an alternative perspective on the organization and relative intensity of vortical flow features. Together, these two visualizations of derived fields provide complementary evidence for interpreting the wake structure behind the half-cylinder.

This case study also highlights HiLSVA's mixed-initiative and self-improving capabilities during scientific analysis. After computing the Q-criterion and visualizing regions with $Q>0$, HiLSVA summarizes this successful workflow through self-reflection and stores it as reusable knowledge. When the user later requests the $\lambda_2$ analysis, the system retrieves this prior experience on vortex identification to guide the subsequent workflow, reducing the need for the user to specify each intermediate step manually. After both visualizations are generated, the user can compare their results and decide how to proceed.

Finally, after obtaining the Q-criterion and $\lambda_2$ visualizations for the $Re=640$ case, the user asks the system to extend the same analysis to the other data with different Reynolds numbers. HiLSVA can therefore apply these computations in batch, although we do not show the additional results here due to space limitations. Overall, this example illustrates how HiLSVA supports \hot{scientific analysis} by going beyond direct visualization of raw data and enabling users to derive, validate, and compare scientifically meaningful flow features. \hot{As a mixed-initiative system, HiLSVA supports human-guided scientific analysis rather than autonomous scientific discovery. Assessing whether such workflows lead to discoveries requires longitudinal studies with domain scientists and remains an important direction for future work.}

\vspace{-0.05in}
\subsection{User Study}

{\bf Experiment setup.}
We conducted a controlled user study with 12 participants under the University's IRB protocol. To examine how HiLSVA supports users with different mental models and prior experience, we recruited participants across three expertise levels: three SciVis experts, four domain scientists with familiarity with scientific datasets but limited experience with visualization tools, and five novices with neither domain nor visualization expertise. Each participant used HiLSVA on a workstation equipped with an NVIDIA RTX 5090 GPU and two 32-inch displays at 3840$\times$2160 resolution. All participants completed the full study and received \$30 compensation. \hot{Two authors administered the study sessions, while all tasks were completed independently by the participants. Throughout the study, HiLSVA logged user-agent interactions and workflow plans and generated visualizations for subsequent quantitative and qualitative analyses.}


To evaluate the effect of human involvement, we studied three autonomy configurations with increasing degrees of mixed initiative: (1) full-autonomous mode, in which the system executed with minimal user intervention; (2) half-autonomous mode without LTT, in which participants could provide guidance but the self-improving component was disabled; and (3) mixed-initiative mode with LTT enabled, in which the system could retrieve prior knowledge, reflect on uncertainty, and proactively solicit user feedback.
\hot{Because these modes vary in both human involvement and the availability of LTT, our study does not isolate the independent effect of LTT. Since LTT depends on user feedback to accumulate and retrieve knowledge, a controlled ablation would also require standardizing user interactions, which we leave to future work.}

For the first three tasks (hurricane, foot, and tornado), each participant was assigned a different autonomy mode for each case, ensuring that each participant experienced all three modes exactly once. We further balanced assignments across participants so that each mode was explored the same number of times. For the fourth task (combustion), participants were allowed to choose their preferred autonomy setting, enabling us to observe which level of control they naturally favored after gaining experience with the system.
We did not include the half-cylinder case in the user study because interpreting the Q-criterion and $\lambda_2$-based vortex identification requires substantial domain knowledge in flow analysis, which could be difficult for novice participants to understand reliably.

{\bf Tasks and procedure.}
The study followed a four-phase workflow: introduction, exploration, task execution, and post-task survey. In the introduction phase, participants watched a short demonstration of HiLSVA using the entropy field from the Richtmyer–Meshkov instability simulation. They were introduced to the main interface and its interaction mechanisms. In the exploration phase, they were given time to freely interact with the system on the same dataset, ask questions, and become comfortable with the workflow before beginning the recorded task session.

In the task execution phase, participants completed four visualization tasks spanning both basic operations and more complex workflow construction. The hurricane case involves slice-based scalar field visualization, histogram inspection, and contour refinement. The focus of the foot case is on isosurface rendering, external anatomy reference retrieval, and visualization refinement for clearer bone identification. The tornado case examines vector field visualization through iterative streamline exploration, trial selection, rollback, and glyph-based refinement. The combustion case involves time-varying multivariate analysis, including timestep identification, volume rendering, and animation of the \texttt{Y\_OH} field, as well as isosurface animation of the \texttt{mixfrac} field. In all tasks, participants also answered several interpretation questions about the resulting visualizations, allowing us to assess both task completion and understanding of the data.

After completing the interactive tasks, participants filled out a post-study questionnaire covering background, usability, interpretability, \hot{perceived transparency}, and the quality of human–agent collaboration. \hot{The questionnaire was designed as a task-specific instrument to evaluate key design goals of HiLSVA. It captures participants' subjective perceptions of system transparency, rather than providing a comprehensive evaluation of trust calibration.} These survey responses, along with the recorded interaction logs and videos, were used to support our qualitative analysis of how users with varying levels of expertise engaged with HiLSVA across different autonomy settings. The detailed user study protocol is provided in the supplementary material.

\begin{table}[htb]
\centering
\caption{
User study results. Time is reported in minutes. The first three tasks use balanced assignments across autonomy modes (FA: full-auto, HA: half-auto, MI: mixed-initiative). In Task 4 (combustion), participants selected their preferred mode, and we report the chosen mode and its execution time. The last row reports mean$\pm$std.
}
\label{tab:user_study_full}
\vspace{-0.1in}
\begin{adjustbox}{width=\columnwidth}
\begin{tabular}{l l ccccccc c}
\toprule
Pcp. & Expertise 
& \makecell{Exp. \\ Time}
& \makecell{Comp. \\ Time}
& \makecell{FA \\ Time}
& \makecell{HA \\ Time}
& \makecell{MI \\ Time}
& \makecell{Task 4 \\ Time}
& \makecell{Task 4 \\ Mode}
& \makecell{Q\&A \\ Accuracy} \\
\midrule
P1  & expert & 15 & 67 & 13 & 22 & 24 & 8  & MI & 12 / 12 \\
P2  & scientist & 9  & 47 & 6  & 9  & 15 & 17 & MI & 12 / 12 \\
P3  & novice    & 7  & 55 & 18 & 12 & 13 & 12 & HA & 11 / 12 \\
P4  & novice    & 10 & 38 & 7  & 7  & 14 & 10 & HA & 12 / 12 \\
P5  & novice    & 5  & 41 & 10 & 13 & 10 & 8  & FA & 12 / 12 \\
P6  & expert & 4  & 37 & 9  & 11 & 9  & 8  & FA & 12 / 12 \\
P7  & novice    & 14 & 37 & 10 & 10 & 9  & 8  & MI & 11 / 12 \\
P8  & expert & 7  & 51 & 11 & 16 & 14 & 10 & FA & 12 / 12 \\
P9  & scientist & 5  & 27 & 7  & 6  & 8  & 6  & FA & 12 / 12 \\
P10 & scientist & 10 & 54 & 10 & 15 & 13 & 16 & FA & 12 / 12 \\
P11 & scientist & 4  & 51 & 10 & 11 & 18 & 12 & FA & 12 / 12 \\
P12 & novice    & 5  & 36 & 7  & 8  & 15 & 6  & FA & 11 / 12 \\
\midrule
& ---
& 7.92$\pm$3.75 
& 45.08$\pm$11.03 
& 9.83$\pm$3.27 
& 11.67$\pm$4.44 
& 13.50$\pm$4.46 
& 10.08$\pm$3.58 
& --- 
& 11.75$\pm$0.45 / 12 \\
\bottomrule
\end{tabular}
\end{adjustbox}
\vspace{-0.1in}
\end{table}

{\bf Task performance.}
As summarized in Table~\ref{tab:user_study_full}, HiLSVA supported strong task performance across all 12 participants: all participants completed the study, and accuracy on the domain-specific interpretation questions averaged 11.75 out of 12. For the first three tasks, we ensured that each task was explored the same number of times under each autonomy mode, enabling a fair comparison of execution time. Under this controlled setup, execution time was lowest in the full-auto mode ($9.83\pm3.27$ min), followed by half-auto mode ($11.67\pm4.44$ min), and highest in mixed-initiative mode ($13.50\pm4.46$ min). 
\hot{Because each participant experienced all three modes, we tested these differences with a Friedman test, which was not significant ($\chi^2(2)=5.09$, $p=0.078$, Kendall's $W=0.19$~\cite{kendall1939}); Holm-corrected pairwise Wilcoxon tests likewise showed no significant pair, though full-auto versus mix-initiative had a large effect size (Cliff's $\delta=-0.51$~\cite{cliff1993}). We therefore interpret the monotonic increase in mean execution time not as a reliable speed difference, but as a tradeoff between oversight and efficiency, where greater user involvement incurs additional interaction time while enabling verification and intervention.}

\hot{Task performance also did not differ significantly across expertise levels. Grouping participants into SciVis experts ($n=3$), domain experts ($n=4$), and non-experts ($n=5$), Kruskal-Wallis tests showed no significant group differences in task completion time ($H=0.92$, $p=0.63$) or exploration time ($H=0.28$, $p=0.87$), and accuracy on embedded domain-interpretation questions was uniformly high ($100\%$, $100\%$, and $95\%$ for the three groups). HiLSVA thus enables non-experts to achieve expert-comparable performance, and we observed no data hallucinations (fabricated values or non-existent fields) in the logged sessions.}

Participants' autonomy choices in the final task further show that no single interaction style is universally optimal. After experiencing all three autonomy settings in the earlier tasks, participants were allowed to choose their preferred mode for the final combustion task: seven selected full-autonomous mode, two selected half-autonomous mode, and three selected mixed-initiative mode. Many participants preferred greater automation once they became familiar with the system, likely valuing efficiency for a task they felt the agent could handle reliably. At the same time, a notable subset of participants still preferred modes that preserved greater human control.

{\bf Additional results.}
Results on the post-study questionnaire and participant feedback are provided in Appendix~\ref{app:usr}.




%% file: src/limitations.tex
\section{Limitations and Future Work}

While HiLSVA presents a novel way to human-in-the-loop SciVis, several limitations remain. First, reliance on external LLM APIs introduces latency, despite partial mitigation through parallel execution in different virtual environments. Second, although LTT and a dynamic knowledge base enable adaptation, performance remains dependent on the quality of user feedback. In specialized scientific domains, relevant knowledge is often tacit or difficult to express in natural language, limiting effective learning. Incorporating curated domain-specific knowledge could further support complex workflows such as feature extraction.
\hot{We also note that HiLSVA's adaptation is retrieval-based, which accumulates and reuses structured knowledge rather than improving the model, and our study does not isolate its gains from those attributable to increased human involvement. Stronger
self-improvement could come from self-evolving agents that discover new designs~\cite{hu2025adas}, edit their own code~\cite{robeyns2025sica}, or harness-level optimization~\cite{lee2026metaharness}, which we leave to future work.}
\hot{Finally, our user study ($N=12$) is powered to detect large within-subject effects but not fine-grained between-group or interaction effects; expertise-stratified and interaction results are therefore descriptive and indicative. Larger-scale, longitudinal studies with domain scientists are needed to confirm these patterns.}

Future work will extend HiLSVA along three directions. 
\hot{First, while the current implementation focuses on ParaView because it provides a rich set of SciVis capabilities and supports MCP tool calls, generated scripts, and direct GUI manipulation within a shared visualization session, the overall architecture is backend-agnostic. We plan to integrate additional visualization and analysis tools to improve generality, including domains such as molecular dynamics, astrophysics, fluid simulation, and bioimage processing.} 
Second, we aim to enhance learning through better uncertainty modeling, multimodal grounding, and long-term personalization. Prior work highlights the importance of uncertainty-aware interaction in human–agent communication~\cite{bansal2019updates, bansal2024challenges}, as well as the role of multimodal human feedback in disambiguating user intent~\cite{li2024mug, yang2025spark}.
%
\hot{Third, we will expand support for non-expert users through recommendation-based analysis assistance, collaborative multi-user environments, and online deployment. By leveraging prior workflows, domain knowledge, and retrieved examples, HiLSVA could proactively suggest candidate analysis strategies with explanations and expected outcomes, allowing users to select and refine appropriate workflows.}

%% file: src/conclusion.tex
\section{Conclusions}

We have presented HiLSVA, a human-in-the-loop agentic system that repositions SciVis as a collaborative process between humans and LLM agents. Through a mixed-initiative agentic architecture, stepwise provenance tracking, and LTT adaptation, HiLSVA enables transparent, controllable, and interactive workflows that support both automation and human oversight. Our evaluation demonstrates that users at all levels of expertise, including novices, can complete non-trivial SciVis tasks with the system, while mixed-initiative interaction provides a flexible balance between efficiency and control. These findings highlight the importance of designing agentic systems that augment human reasoning rather than replace it. We believe that HiLSVA offers a practical foundation for future human-centered SciVis systems and opens new directions for integrating \hot{transparent, collaborative} AI into complex scientific workflows.

%% file: src/appendix.tex
\newpage
\clearpage

\setcounter{section}{0}
\setcounter{figure}{0}
\setcounter{table}{0}

\section{Additional User Study Results}
\label{app:usr}

\begin{figure}[htb]
\centering
\includegraphics[width=1.0\linewidth]{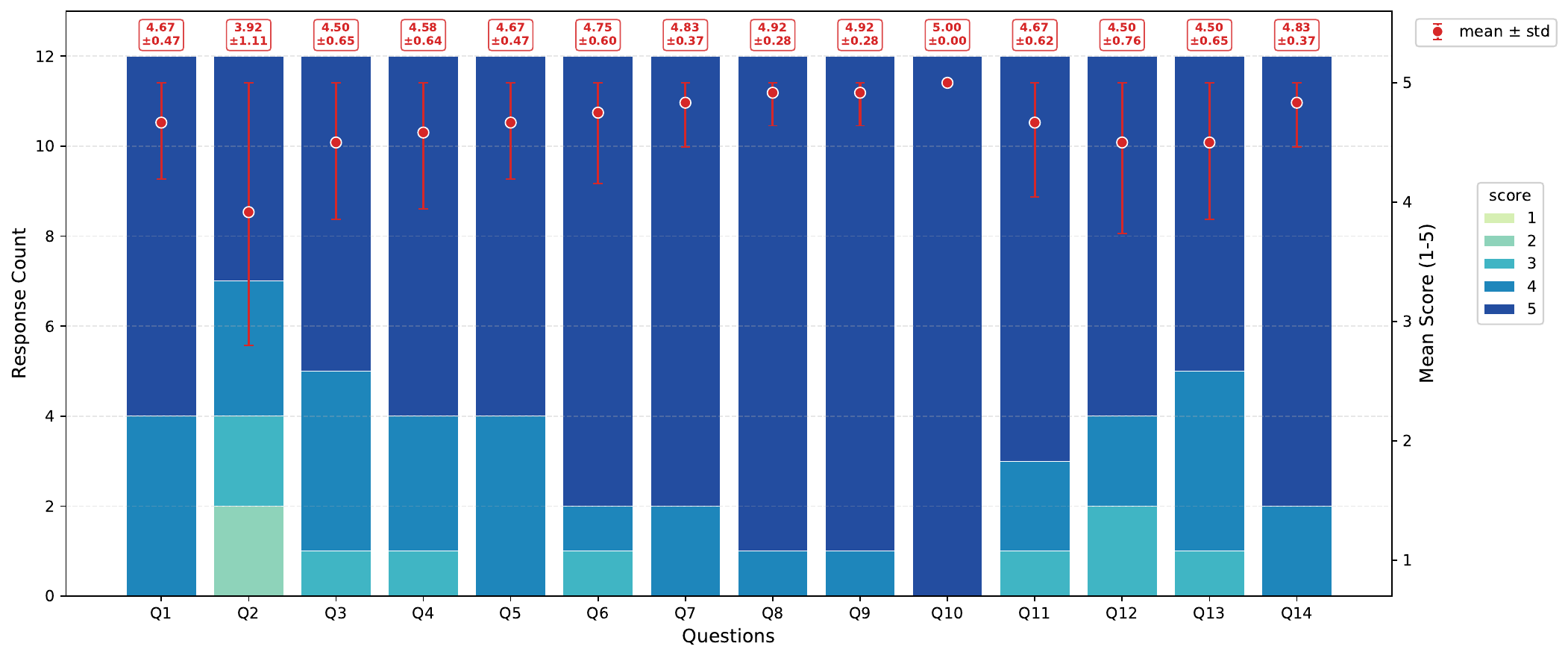}
\vspace{-0.25in}
\caption{
Survey responses to 14 post-study questions, along with mean scores (on a 1–5 scale) and standard deviations.
}
\label{fig:survey}
\end{figure}

{\bf Quantitative results.}
We evaluated the accessibility and perceived effectiveness of HiLSVA through a post-study questionnaire. After completing the tasks, all participants provided ratings on a 1–5 Likert scale for 14 questions, along with qualitative feedback on the system's strengths, challenges, and potential improvements (see the supplementary material). The aggregated results are shown in Figure~\ref{fig:survey}.

Overall, participants reported highly positive experiences with the system, with an average score of 4.66 across all questions.
\hot{Collapsing the two expert groups ($n=7$), experts rated the system slightly lower than non-experts ($4.51$ vs.\ $4.87$; Mann-Whitney $p=0.080$, Cliff's
$\delta=-0.63$), consistent with non-experts benefiting more from agent assistance while expert users judged it more critically.}
Ratings of efficiency and overall effectiveness were consistently strong, with Q1 (task efficiency) receiving 4.67$\pm$0.47. The system's integration and learnability were also well received, with Q3 (system integration) at 4.50$\pm$0.65 and Q4 (ease of learning) at 4.58$\pm$0.64, indicating that participants quickly understood and adopted the workflow.

Several questions focused on human-agent interaction and control. Participants expressed a strong sense of control over the workflow, with Q6 (ability to review and approve actions) rated at 4.75$\pm$0.60 and Q7 (ability to adjust autonomy levels) at 4.83$\pm$0.37. Transparency-related aspects received some of the highest scores in the study, including Q8 (clarity of system feedback, 4.92$\pm$0.28), Q9 (usefulness of stepwise provenance for understanding system behavior, 4.92$\pm$0.28), and Q10 (ability to revisit previous steps, 5.00$\pm$0.00). These results suggest that the system's stepwise planning and provenance-tracking mechanisms effectively supported users' understanding and \hot{oversight}.

The mixed-initiative design was also positively evaluated. Participants rated the combination of natural-language interaction and direct manipulation highly (Q11, 4.67$\pm$0.62) and agreed that the system could recognize uncertainty and appropriately request guidance (Q12, 4.50$\pm$0.76). In addition, Q13 (benefit from accumulated knowledge) received 4.50$\pm$0.65, indicating that the LTT mechanism was noticeable and useful during interaction. The overall interface usability was similarly strong, with Q14 rated at 4.83$\pm$0.37.

The lowest score was observed for Q2 (ability to complete tasks without assistance), with 3.92$\pm$1.11. The larger variance suggests differing expectations across participants, particularly between experts and novices. This is expected, as the four tasks involve non-trivial SciVis techniques that are challenging without prior experience. Despite this, all participants completed the tasks with high accuracy when the system provided support. This aligns with HiLSVA's design goal as a collaborative system, where human involvement remains integral while the agent lowers barriers to complex visualization workflows.

{\bf Qualitative feedback.}
Participants across all levels of expertise provided insightful feedback on the strengths and limitations of HiLSVA. A common theme among all groups was the value of \emph{human–agent collaboration}. Several participants highlighted the system's ability to support iterative refinement, with a SciVis expert highlighting ``{\em the way that user and agent can work together to refine the visualization},'' and a domain scientist emphasizing the usefulness of step-by-step planning for first-time users.

Another frequently mentioned strength was the system's \emph{interactivity and controllability}. Domain scientists emphasized that the system is ``{\em controllable for visualization through text description},'' \hot{highlighting the value of expressing visualization intent without requiring detailed knowledge of visualization software or scripting.} Participants also highlighted the mixed-initiative interaction, particularly the agent's ability to request guidance when uncertain, described as ``{\em a very cool feature}.'' In addition, users appreciated the integration of multiple interaction mechanisms, including both code-based execution and direct tool control through MCP. 
\hot{They also valued the system's autonomous capabilities, particularly its ability to integrate external tools and information sources within a unified workflow, reducing context switching and supporting iterative refinement.}

Despite these strengths, participants reported challenges related to \emph{latency and system responsiveness}. Multiple users noted that the system could be slow sometimes, with comments such as ``{\em LLM takes too long to respond}'' and ``{\em the latency issue makes the visualization procedure pretty long}.'' In addition, several participants mentioned issues with system stability, including GUI inconsistencies and occasional crashes.

Feedback also revealed opportunities for improving \emph{information presentation}. Some expert users found the system outputs overly verbose, noting that ``{\em output always contains too much information without highlighting},'' and suggested mechanisms to prioritize critical information. Novice users, on the other hand, expressed a need for more guidance, including clearer explanations of visualization concepts and domain-specific terminology.
\hot{While tasks that require greater analysis and domain knowledge can benefit from stronger human oversight, novice users may still prefer higher autonomy levels because they lack the expertise to guide the workflow.}

Finally, participants suggested improvements to \emph{workflow efficiency and usability}, including better session management, voice interaction, and more expert-curated default settings. In particular, several novice users requested presets derived from expert knowledge to effectively create visualizations without requiring domain expertise.

\hot{These observations reveal several practical limitations and opportunities for improvement. Latency was primarily associated with planning and external LLM response times. GUI inconsistencies were caused by known synchronization limitations in ParaView that would require backend-level modifications to address fully. Conflicting operations typically caused occasional crashes, although the containerized architecture enables straightforward session recovery. User feedback on verbose outputs motivated a redesign that suppresses intermediate reasoning and tool-calling traces in favor of concise analytical results. Finally, the differing needs of novice and expert users suggest opportunities for adaptive guidance, recommendations, and expert-derived presets that better support complex analytical tasks.}